\documentclass[sigconf]{acmart}

\usepackage{booktabs} 

\usepackage{hyperref}
\usepackage{graphicx}
\usepackage{url}

\usepackage[english]{babel}

\usepackage{amsthm}
\usepackage{todonotes}
\usepackage{amsmath}
\usepackage{amssymb}
\usepackage{url}
\usepackage{listings}
\usepackage{balance}
\usepackage{epigraph}
\usepackage{fixltx2e}
\usepackage{graphicx}
\usepackage{algorithm}
\usepackage{algpseudocode}
\usepackage{caption}
\usepackage{color}
\usepackage{hyperref}
\usepackage{breakurl}
\usepackage[titletoc,title]{appendix}
\usepackage[inline]{enumitem}
\usepackage{eurosym}

\usepackage[T1]{fontenc}
\usepackage[utf8]{inputenc}

\usepackage[labelformat=simple]{subcaption}

\hypersetup{
  colorlinks=true,
  citebordercolor=cyan,
  filebordercolor=red,
  linkbordercolor=blue,
  citecolor=cyan,
  linkcolor=red,
  urlcolor=blue}

\makeatletter
\let\OldStatex\Statex
\renewcommand{\Statex}[1][3]{%
  \setlength\@tempdima{\algorithmicindent}%
  \OldStatex\hskip\dimexpr#1\@tempdima\relax}
\makeatother

\everymath{\displaystyle}

\DeclareGraphicsExtensions{.png,.pdf}

\begin{document}

\copyrightyear{2017}
\acmYear{2017}
\setcopyright{acmlicensed}
\acmConference{PPDP'17}{October 9--11, 2017}{Namur, Belgium}\acmPrice{15.00}\acmDOI{10.1145/3131851.3131862}
\acmISBN{978-1-4503-5291-8/17/10}

\title{Practical Evaluation of the \\ Lasp Programming Model at Large Scale}
\subtitle{An Experience Report}

\author{Christopher S. Meiklejohn}
\affiliation{%
  \institution{Universit\'e catholique de Louvain}
  \city{Louvain-la-Neuve}
  \country{Belgium}
}

\author{Vitor Enes}
\affiliation{%
  \institution{HASLab / INESC TEC}
  \institution{Universidade do Minho}
  \city{Braga}
  \country{Portugal}
}

\author{Junghun Yoo}
\affiliation{%
  \institution{University of Oxford}
  \city{Oxford}
  \country{United Kingdom}
}

\author{Carlos Baquero}
\affiliation{%
  \institution{HASLab / INESC TEC}
  \institution{Universidade do Minho}
  \city{Braga}
  \country{Portugal}
}

\author{Peter Van Roy}
\affiliation{%
  \institution{Universit\'e catholique de Louvain}
  \city{Louvain-la-Neuve}
  \country{Belgium}
}

\author{Annette Bieniusa}
\affiliation{%
  \institution{Technische Universit{\"a}t Kaiserslautern}
  \city{Kaiserslautern}
  \country{Germany}
}

\renewcommand{\shortauthors}{C. Meiklejohn et al.}

\begin{abstract}
Programming models for building large-scale distributed applications assist the developer in reasoning about consistency and distribution.
However, many of the programming models for weak consistency, which promise the largest scalability gains, have little in the way of evaluation to demonstrate the promised scalability.
We present an experience report on the implementation and large-scale evaluation of one of these models, Lasp, originally presented at PPDP `15, which provides a declarative, functional programming style for distributed applications.
We demonstrate the scalability of Lasp's prototype runtime implementation up to 1024 nodes
in the Amazon cloud computing environment.
It achieves high scalability by uniquely combining hybrid gossip with a programming model
based on convergent computation.
We report on the engineering challenges of this implementation and its evaluation,
specifically related to operating research prototypes in a production cloud environment.
\end{abstract}

\maketitle

\section{Introduction}


Once a specialized field for applications that required large data sets, large-scale distributed applications have become commonplace in our globalized society.
Regardless of whether you are developing a rich-web application or a native mobile application, managing distributed data is challenging.
For simplicity, developers today typically resort to using a single database that provides a form of strong\footnote{For instance, linearizability, where a value follows the real-time order of updates.} consistency.
In essence, the database serves as shared memory for the clients in the system.


A single database is an obvious bottleneck as it introduces a serialization point for all operations; this restricts the possible throughput of the system.
As developers strive to provide a near-native experience where operations appear to happen immediately, and since not all clients can be geographically located close to the database, application performance can suffer as users move farther from the database; or worse, when clients can't communicate with the database at all because they are offline.
To provide good user experience, including high availability and low latency, developers are forced to integrate replication in the system design.


Systems that favor weak consistency scale better: data items can be locally replicated, locally mutated by the application, and their state can be disseminated asynchronously, outside of the critical path.  Weak consistency allows applications to continue to operate while offline.  While these systems provide for high scalability and high performance, programming with weak consistency can be a challenge for the application developer as updates to data items have no guarantee on update visibility or update order.  Concurrency poses an additional problem, as updates happening concurrently at different replicas may be conflicting.


Numerous systems and programming models\cite{Sivaramakrishnan:2015:DPO:2737924.2737981, conway2012logic, terry1995managing, meiklejohn2015lasp, alvaro2011consistency, kuperjoining, burckhardt2015global, myter2016now} have been proposed for working with weak consistency, however few have seen adoption.  Many of the systems have sound theoretical foundations, but few perform evaluations at scale to demonstrate the benefits in practice.  We believe that the lack of these results comes from the difficulty in the required  infrastructure for large-scale experiments, and the challenges in engineering an implementation of a theoretical model using existing software languages and libraries.

In this paper, we discuss the practical issues encountered when evaluating
one of these programming models, Lasp~\cite{lasp-implementation, lasp-documentation}, originally presented at PPDP `15.
Lasp is designed using a holistic approach where the programming model was co-designed
with its runtime system to ensure scalability.
We examine the challenges of engineering an implementation capable of scaling to a large number of nodes running in a public cloud environment, using a real world application scenario.
Further, we report on the engineering challenges of demonstrating the scalability of the Lasp model.
Our experience report substantiates that empirically validating scalability is non-trivial, regardless of the programming model.

\section{Advertisement Counter}

Lasp was invented to ease the development of distributed applications with weak consistency.  The advertisement counter scenario from Rovio Entertainment, creator of Angry Birds, is an ideal fit for Lasp.  This application counts the total number of times each advertisement is displayed on all client mobile phones, up to a given threshold for each.
The application has the following properties:
\begin{itemize}
\item \textbf{Replicated data.} 	Data is fully replicated to every client in the system.  This replicated data is under high contention by each client in the system.
\item \textbf{High scalability.} Clients resemble individual mobile phone instances of the application, so the application should scale up to millions of clients.
\item \textbf{High availability.} Clients need to continue operation when disconnected as mobile phones frequently have periods of signal loss (offline operation).
\end{itemize}

As part of the large-scale evaluation done in the SyncFree project,
and following the personal curiosity of the developers, we decided to invest resources in using industrial-strength engineering techniques to evaluate the scalability of this application running in a real world production cloud environment.

\subsection{Lasp}

Lasp~\cite{meiklejohn2015lasp} is a programming model that allows developers to write applications with Conflict-Free Replicated Data Types (CRDTs)~\cite{shapiro2011comprehensive, DBLP:journals/corr/AlmeidaSB16}.  CRDTs are abstract data types, designed for use in concurrent and distributed programming, that have a binary merge operation to join any two replicas of a single CRDT.  Under concurrent modification without coordination, different replicas of a single CRDT may diverge; the merge operation supports value convergence by ensuring that given enough communication, all replicas, without coordination, will converge to a single deterministic value regardless of the order that data is received and merged.

Historically, before CRDTs were introduced, ad-hoc merge functions were used, often with few formal guarantees.
Later, after their development, programmers who wanted to use CRDTs in their applications would have two choices: either, using a single CRDT from existing literature to store application state, fitting their problem to an existing data structure; or, building a custom CRDT that fits their application domain, which requires to ensure that the merge operation is both deterministic and convergent.

Lasp improves this choice in two ways:

\begin{itemize}
\item \textbf{Composition.} Lasp provides set-theoretic and functional combinators for composing CRDTs into larger CRDTs.
\item \textbf{Monotonic conditional.} Lasp introduces a conditional operation that allows the execution of application logic based on monotonic conditions\footnote{Monotonicity implies that once a condition becomes true, it remains true; a monotonicity check can be done without distributed coordination.} on CRDTs.
\end{itemize}

These two concepts allow Lasp applications to be both transparently and arbitrarily distributed across a set of nodes without altering application behavior.  For brevity, the reader is referred to~\cite{meiklejohn2015lasp} for a full treatment of the Lasp semantics.

The advertisement counter uses two data structures from Lasp: the Add-Wins Set CRDT\footnote{a.k.a. Observed-Remove Set}, where elements can be arbitrarily removed and inserted without coordination and under concurrent add and remove operations the add will `win'; and the Grow-Only Counter CRDT, which models a counter that only increments.

\subsection{Overview}

\begin{figure}[h]
\begin{center}
\noindent\includegraphics[scale=0.25]{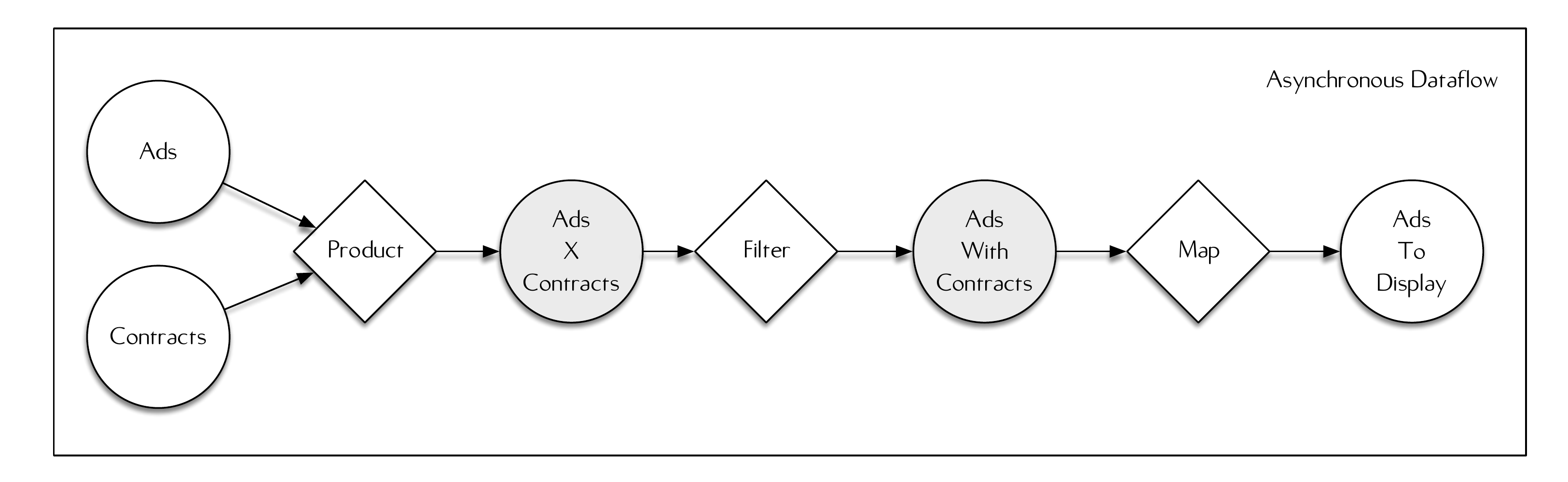}
\end{center}
\caption{Asynchronous dataflow computation in Lasp that derives the set of displayable advertisements.}
\label{fig:advertisement-counter-async-dataflow}
\end{figure}

The design of the advertisement counter is roughly broken into three components.
\begin{itemize}
\item \textbf{Initialization.} When the advertisement counter application is first initialized, we first create Grow-Only Counters for each unique advertisement we want to track impressions for, and we then insert references to them into an initial Add-Wins Set of advertisements.  
\item \textbf{Selection of displayable advertisements.}  We define a dataflow computation in Lasp that will derive an Add-Wins Set of advertisements to display to the clients based on advertisements that have valid ``contracts'': records that represent that an advertisement is allowed to be displayed at the current time
(Figure~\ref{fig:advertisement-counter-async-dataflow}).
\item \textbf{Enforcing invariants.}	  Since clients increment each advertisement counter as advertisement impressions occur, when the target number of impressions is reached both the client and the server will fire a trigger to remove the advertisement counter from the set of advertisements, to prevent the advertisement from being further displayed.  This can be done without coordination through the use of the Add-Wins Set.  
\end{itemize}


The advertisement counter has two important design choices, which makes its implementation in Lasp ideal.
\begin{itemize}
\item \textbf{Offline support.} As Angry Birds is a mobile application, there will be periods without connectivity.  During this time, advertisements should still be displayable.
\item \textbf{Lower-bound invariant.} Advertisements need to be displayed a minimum number of times; additional impressions are not problematic.  This is a monotonic condition: once the condition is true, it remains true.
\end{itemize}

\subsection{Implementation}

The advertisement counter is broken into two components that work in concert.  Both components track a single replica of a set of identifiers of displayable advertisements, and for each identifier a replica of an advertisement counter that tracks the total number of times the advertisement has been displayed to the user.  Each node in our experiment runs either a single client or server process.

\begin{itemize}
\item{\textbf{Server processes.}} One or more server processes, each responsible for propagating their state to clients and disabling advertisements that have been displayed a minimum number of times by monotonically removing them from the set of displayable advertisements.
\item{\textbf{Client processes.}} Many client processes that periodically propagate their state with other nodes, and increment their counter replicas based on a synthetic workload.  
\end{itemize}

The prototype implementation of the Lasp programming model is built in the Erlang programming language and exposed to the user as an application library.

The fully instrumented Lasp advertisement counter client is implemented in 276 lines of Erlang code, and the fully instrumented advertisement counter server is 333 lines of Erlang code.  Around 50$\%$ of this code is for instrumentation and  orchestration, to ensure we can perform a full analysis of the application during experimentation.  The Lasp runtime system takes care of cluster maintenance, data synchronization and storage, which are done manually in the previous approaches (ad-hoc merge or custom CRDT design).

\section{System Architecture}

To perform a real world evaluation of the advertisement counter, we implemented an efficient, scalable runtime system for Lasp.  Lasp's runtime system is a highly-scalable eventually consistent data store with two different dissemination mechanisms (state-based vs. delta-based) and two different cluster topologies (datacenter vs. hybrid gossip).  Lasp's programming model, presented in~\cite{meiklejohn2015lasp}, sits above the data store and exposes a programming interface.

Datacenter Lasp~\cite{meiklejohn2015lasp} operates using a structured overlay network.   Hybrid Gossip Lasp~\cite{meiklejohn2015selective} uses an unstructured overlay network, and by design should achieve greater scalability and provide better fault-tolerance~\cite{rodrigues2010peer}.

\subsection{Datacenter Lasp}

Datacenter Lasp refers to the prototype implementation of the runtime system presented with the programming model, at this conference two years ago~\cite{meiklejohn2015lasp}.

In Datacenter Lasp, all CRDT state is both partitioned and replicated across several datacenter nodes. Client processes communicate directly with server processes that are running on datacenter nodes; client processes do not communicate amongst each other.  Replication is used across datacenter nodes for fault tolerance, and partitioning/sharding is used for horizontal scalability: this is achieved through the use of consistent hashing and hash-space partitioning.  In our experiments this is simplified and there is no partitioning, since the data set for our experiments never exceeds a single datacenter node's available capacity.

\subsection{Hybrid Gossip Lasp}

Hybrid Gossip Lasp is inspired by two Hybrid Gossip protocols, HyParView~\cite{leitao2007hyparview}, and Plumtree~\cite{leitao2007epidemic}.  In Hybrid Gossip Lasp, nodes are assembled in a peer-to-peer topology, where client processes can communicate either with server processes running on datacenter nodes or client processes.  State is delivered transitively through other processes in the system: there is no need to communicate directly with a server process running on a datacenter node.


Hybrid Gossip Lasp uses a membership protocol heavily inspired by HyParView, to compute an overlay network containing all of the members in the cluster.  The notable differences between the HyParView protocol and our membership protocol were the results of adapting the theoretical treatment in the HyParView paper to an actual implementation that was used for this experiment.

Specifically, the original HyParView protocol was evaluated in a low-churn environment, whereas our environment has much higher churn.  {\em Churn} is defined as rate of node turnover, i.e., percentage of nodes leaving and being replaced by new nodes, per time unit.  The higher churn in our environment was a byproduct of attempting to reduce experimentation time to save costs when operating large clusters: this allowed experiments that would normally take hours for cluster deployment and operations to be reduced to fractional hours at significant cost savings.  For details on the modifications to the protocol, the reader is referred to~\cite{meiklejohn2017loquat}.

\subsection{Dissemination Protocols}

The system supports two data dissemination protocols.

\begin{itemize}
\item{\textbf{State-based.}} Objects are locally updated through mutators that inflate the state.
Objects are periodically sent to
peers that merge the received object with their local state. 
\item{\textbf{Delta-based.}} Objects are locally updated by merging the state with the result of $\delta$-mutators \cite{DBLP:journals/corr/AlmeidaSB16}, called deltas, that compactly represent changed portions of state. These deltas are buffered locally and sent to each local peer in every propagation interval.

\end{itemize}

\section{Engineering Scale}

The Lasp semantics ensures that the runtime system is correct in theory for arbitrary
distribution of the computation.
However, engineering a scalable real-world system requires
a significant amount of sophisticated tooling
to ensure scalability both for deployment and
for observability during execution.
Near the end of the SyncFree project, we designed an experiment
with the goal of scaling to 10\,000 nodes.
We finally achieved a scale of 1024 nodes at a total cloud computing cost of about \euro 9000.

\subsection{Experiment Configuration}

For the purposes of the experiment, we used a total of 70 m3.2xlarge instances in the Amazon EC2 cloud computing environment, within the same region and availability zone.  We used the Apache Mesos~\cite{hindman2011mesos} cluster computing framework to subdivide each of these machines into smaller, fully-isolated machines using cgroups.  Each virtual machine, representing a single Lasp node, communicated with other nodes in the cluster using TCP, and given the uniform deployment across all of the allocated instances, had varying latencies to other nodes in the system depending on their physical location.

When subdividing resources for the experiment, we allocated each server task 4\,GB of memory
with 2 virtual CPUs, and each client task 1\,GB of memory, with 0.5 virtual CPUs.
Here a {\em task} is a logical unit of computation
that is executed on one virtual machine.
We consider that these numbers vastly underrepresent the capabilities of modern mobile devices in widespread deployment today and therefore will lead to conservative results in the evaluation.  We allocate more resources to servers, specifically in Datacenter Lasp mode, as servers are required to maintain connections to more nodes in the system; the advertisement counter does not require more resources between Datacenter and Hybrid Gossip modes.

\subsection{Experimental Workflow}

As running experiments in an unsimulated cloud environment can be challenging due to the inherent nondeterminism across different executions of the same experiment, we created a workflow targeted at reducing nondeterminism by controlling the experiments' setup and teardown procedures with detailed instrumentation for post-experimental analysis.   We describe that workflow below.

\begin{itemize}

\item{\textbf{Bootstrapping.}}  Initially, all of the server and client processes are bootstrapped and joined into a single cluster.
The experiment does not begin until we ensure that all of the nodes in the system are connected and the connection graph forms a single connected component.  Each node should be reachable by every other node in the system, either directly as a local neighbor, or indirectly via multi-hop. During this process, the system creates advertisement counters and the set of displayable ads.
\item{\textbf{Simulation.}} Once we ensure the cluster is connected, each node starts collecting metrics and generating its own workload that randomly selects a counter to increment based on the set of displayable advertisements every predefined impression interval.  Periodically, each process propagates local replicas with neighbor processes.  
It should be noted that each client has its own workload generator: using a centralized harness for running the experiment introduces coordination, which reduces the scalability of the system.
\item{\textbf{Convergence.}} As each of the experiments has a controlled number of events that will be generated based on the number of clients participating in the system, the experiment continues to run until each node has observed the effects of all events: we refer to this process as convergence.  
\item{\textbf{Metrics aggregation and archival.}} Once convergence is reached, the experiment is complete.  Each node, upon observing convergence begins uploading metrics recorded during the experiment to a central location: these logs are used for analysis of the runtime system.  Once this process is complete, the experiment harness waits for the system to fully teardown the cluster before starting a subsequent run, to prevent state leakage between runs when reusing the same hardware to reduce costs.
\end{itemize}

\subsection{Experimental Infrastructure}

Evaluation of a large-scale distributed programming model is difficult.  This is due to failures in the underlying frameworks that are used to provide mechanisms for deployment and operations, and because of inadequate tools required to observe the system during execution to ensure it is operating properly.

\subsubsection{Apache Mesos}

While experimentation shows Lasp scalability to 1024 nodes, we do not believe that this number is a firm upper limit.  When attempting to run experiments with 2048 nodes we quickly ran into problems with the Apache Mesos cloud computing framework.  One issue is that when attempting to bootstrap a cluster containing 70 instances too quickly, instances become disconnected and need to be manually reprovisioned.  This required a slower cluster deployment where a cluster would be scaled from 35 instances, first to 50 instances, and then to 70 instances.  As the 2048 experiment required 140 m3.2xlarge instances to operate, cluster deployment would take significantly longer.

When attempting to launch 2048 tasks in Mesos (with a single task representing a single application node),
instances would become overloaded quickly and fail to respond to heartbeat messages: this triggered these instances being marked as offline by Mesos and the tasks orphaned.  This would require restarting the experiment and reallocating the cluster to account for the lost tasks.

\subsubsection{Sprinter}

Once tasks were launched by Apache Mesos, we needed a mechanism for client processes to discover other client processes in the system and connect to them.

Therefore, we built an open source service discovery library called Sprinter that was used to fetch a list of running tasks from the Mesos framework, Marathon, and supply them to the system as targets to connect to.  Sprinter also performs the following functions:

\begin{itemize}
\item \textbf{Graph analysis for connectedness.} Each node uploads its local membership view to Amazon S3.  The first, lexicographically ordered, server periodically pulls this membership information and builds a local graph that is analyzed to determine if the graph contains all clients, and that the connection graph forms a single connected component.
\item \textbf{Delay experiment for connectedness.} Based on graph analysis, the experiment's start is delayed until the connection graph forms a single connected component.
\item \textbf{Periodic reconnection if isolated.} If a node becomes isolated from the cluster, it will rejoin the cluster, using the information provided by Marathon.
\end{itemize}

To assist in operator debugging of the experiments, a graphical tool was built to visualize the graph information from Sprinter along with extensive logging to the server node with information about cluster conditions.

\subsubsection{Partisan} Distributed Erlang has known scalability problems when operated in the range of 50 or more nodes as it tracks full membership information in the cluster at each node and maintains full connectivity between nodes using a single TCP connection that is used for both data transmission and heartbeat messages.  Single connections are problematic because of head-of-line blocking when large messages are transmitted.

We knew that for the experiment to scale we would need: (1) to move away from Distributed Erlang, (2) to configure network topologies for both Datacenter Lasp and Hybrid Gossip Lasp in a single specification, and (3) to specify configurations at runtime without having to modify application code.  To do this we built Partisan, an open source Erlang library that provides an alternative communication layer that eschews the use of Distributed Erlang.  Partisan supports multiple network configurations and topologies: a client-server star topology, a full connectivity topology mirroring Distributed Erlang's, a static topology where per-node membership is explicitly maintained, and a random unstructured overlay membership protocol inspired by the HyParView membership protocol.

\subsubsection{Workflow CRDT (W-CRDT)}

In our experiments, a central task could not be used to orchestrate the execution: early experiments demonstrated that the central task quickly became a bottleneck and slowed down execution to the speed of the central task.  Therefore, we eliminated the central task.

However, without a central task performing orchestration, it becomes more difficult to control when nodes should perform certain actions.  For example, after event generation is complete, we should wait for convergence before proceeding to metrics aggregation.  Therefore, we needed a mechanism for asynchronously controlling the workflow of the application scenario.

We devised a novel data structure, called the \textit{Workflow-CRDT} (W-CRDT), that is disseminated between nodes for controlling when certain actions should take place.  This object is not instrumented by our runtime or included in any of the application logging, to prevent the structure itself from influencing the results of the experiment.  The W-CRDT is a sequence of Grow-Only Map CRDTs, where each map is a function from opaque node identifiers to booleans.  The sequence is implemented with the recursive Pair CRDT
(similar to a recursive list type).
The W-CRDT operates as follows:
\begin{itemize}
\item \textbf{Per node flag.} Each node's portion of a task to be completed is modeled as a flag; each node toggles its flag when it has completed its work.
\item \textbf{Tasks as grow-only maps.} Each task that needs to be performed is represented by one grow-only map. When all the map's flags are true, the task is considered as complete.  This corresponds to a barrier synchronization.
\item \textbf{Sequential composition of tasks.} Each task can be sequenced with another task.  A task starts when its preceding task has completed.
\item \textbf{Workflow completion.} The workflow is considered complete when all of the tasks that make up the sequential composition are complete.
\end{itemize}


The W-CRDT is used to model the following sequential workflow in each experiment. 
\begin{itemize}
\item \textbf{Perform event generation.} Once event generation is complete, nodes mark event generation complete.
\item \textbf{Blocking for convergence.} Once convergence is reached, nodes mark convergence complete.
\item \textbf{Log aggregation.} Once convergence is reached, nodes begin uploading their logs to a central location and mark log aggregation complete.
\item \textbf{Shutdown.} Shutdown once log aggregation is complete.\end{itemize}

\section{Evaluation}

For Datacenter Lasp, we ran experiments using state-based dissemination, with a single server, and 32, 64, 128, 256 clients, forming with the server a star graph topology. For Hybrid Gossip Lasp, we ran experiments using both dissemination strategies, with a single server, and 32, 64, 128, 256, 512, and 1024 clients.  

Each experiment was run twice, with the advertisement impression interval fixed at 10 seconds and the propagation interval at 5 seconds. The total number of impressions was configured to ensure that, in all executions, the experiment ran for 30 minutes.

\begin{figure}[h]
\begin{center}
\noindent\includegraphics[scale=0.6]{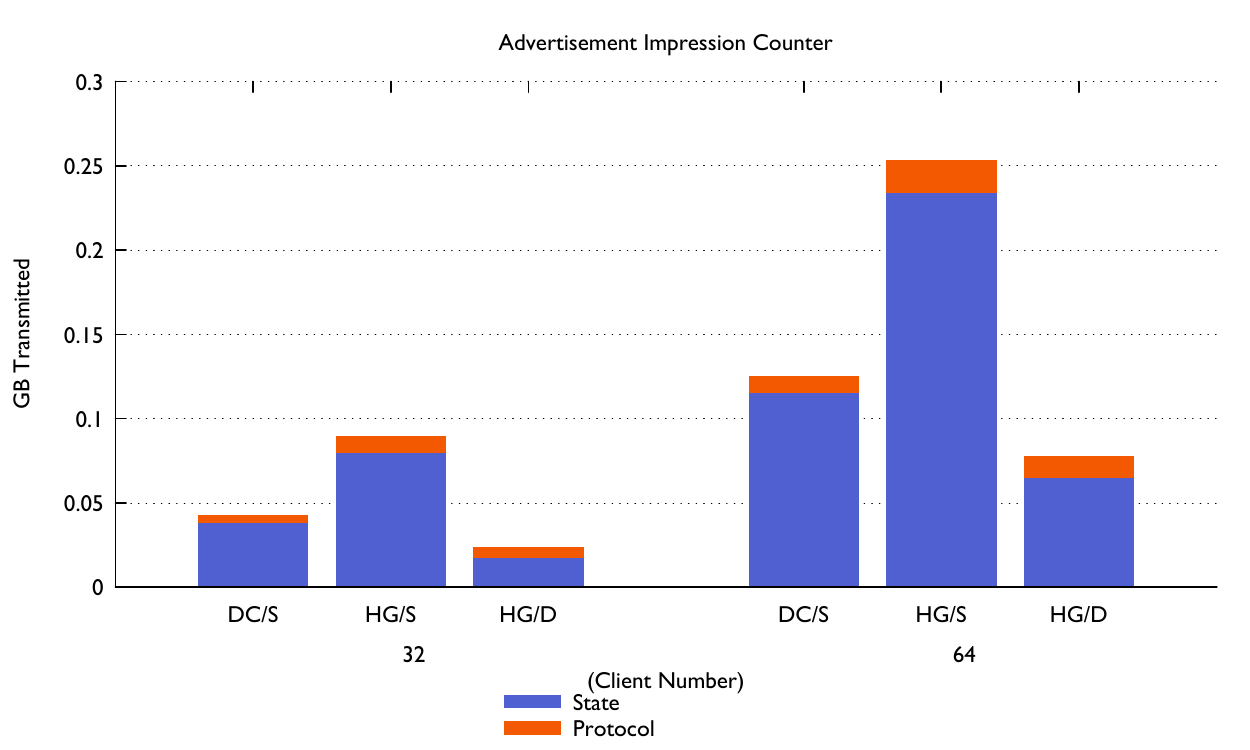}
\end{center}
\caption{Comparison of state- and delta-based dissemination in both Datacenter and Hybrid Gossip Lasp with 32/64 clients.}
\label{fig:transmission-modes}
\end{figure}

\begin{figure}[h]
\begin{center}
\noindent\includegraphics[scale=0.6]{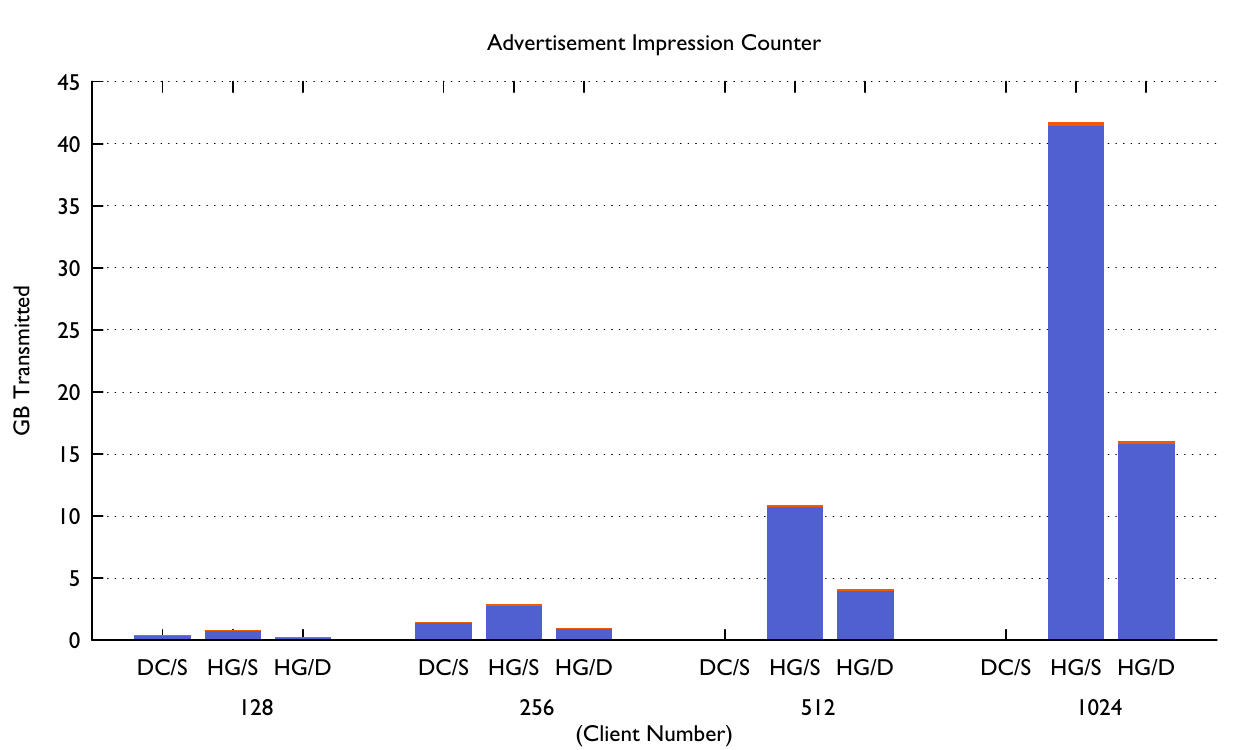}
\end{center}
\caption{Comparison of state- and delta-based dissemination in both Datacenter and Hybrid Gossip Lasp with Datacenter Lasp $\leq$ 256 clients (limited in scalability) and Hybrid Gossip Lasp $\leq$ 1024 clients.}
\label{fig:transmission-scale}
\end{figure}

Figure~\ref{fig:transmission-modes} and Figure~\ref{fig:transmission-scale} evaluate three different operational modes for Lasp, examining the state transmission for the duration of the experiment.  Two Hybrid Gossip dissemination strategies, state-based and delta-based, are evaluated using a single overlay generated by the HyParView protocol.  We also evaluated Datacenter Lasp, where clients propagate changes to the server using a state-based dissemination strategy.  We did not evaluate delta-based for Datacenter Lasp, as it is unrealistic to believe that the server could buffer all changes in the system. In this evaluation, we scale up to 256 client processes: this is the largest number of client processes a single server could support in Datacenter Lasp.  Hybrid Gossip scaled to 1024 nodes, before we ran into issues with Apache Mesos.

Datacenter Lasp performs the best in terms of state transmission when compared to Hybrid Gossip Lasp using the same dissemination strategy.  This results from Datacenter Lasp have no redundancy at all: the star topology has a single point of failure that is used for communication between all nodes in the system.  Delta-based dissemination demonstrates a clear advantage for Hybrid Gossip Lasp where redundancy is required to keep the system operating: state transmission can be reduced without sacrificing the fault-tolerance properties of the underlying overlay network. In terms of protocol transmission in Hybrid Gossip Lasp, delta-based dissemination performs better than state-based, even though it is a more complex protocol: in delta-based dissemination a process can track which updates have been seen by its neighbor processes and it will not disseminate an unchanged object, while in state-based dissemination an object is always propagated.

%

Our experiments confirm several design considerations made in Lasp.  First, as demonstrated by the graphs, in the Datacenter Lasp model the transmission cost is reduced as there is no redundancy in messaging and subsequently no fault-tolerance.  In this model, because of communication through a datacenter node, an update takes two hops to reach all clients in the system.  However, this model has limited scalability because a centralized point, which could be partitioned and replicated across multiple servers, is used as a coordination point for all clients.

Hybrid Gossip Lasp adds additional redundancy by constructing a random overlay network using the HyParView protocol and gossiping state to local peers.  This model has additional cost, but provides fault-tolerance through redundancy. In the worst case, an update will be observed by all nodes $V$ after $\log \left| V \right|$ propagation intervals, since in this topology the diameter is logarithmic on the number of nodes.


%

\section{Conclusion}

Designing new programming models for building large-scale distributed applications requires not only a solid theoretical design, but a well-engineered solution to demonstrate that the system can scale as advertised.  Specifically, large-scale evaluations are plagued by the following problems.
\begin{itemize}
\item \textbf{Existing tooling can be problematic.} Existing infrastructure, frameworks, and languages can be treacherous as they can reduce the scalability of the system because of their design choices.  
\item \textbf{Visualizations are invaluable.} Visualizations assist in debugging the system in real time. 
\item \textbf{Achieving reproducibility is non-trivial.}  Clouds provide high-level abstractions over machines, removing visibility into server location and isolation which makes controlled experiments difficult.
\item \textbf{Performance can fluctuate.} Virtual machine placement and migration, compounded by a language VM layer, are factors that make performance measurement unpredictable. Cost considerations also limit the statistical smoothing possible by running multiple experiments.
\item \textbf{Evaluations are expensive.}  To provide a real world evaluation,
significant funding is required for the infrastructure resources
and significant time is required for developing deployment tools and for debugging experiments.
\end{itemize}

Lasp's scalable design was achieved by taking a holistic approach: both the runtime system and programming model were designed to accommodate one another in a way that allows scalability.  However, the effort required to demonstrate Lasp as both scalable and practical remained a non-trivial challenge.

{\small
\paragraph{Acknowledgements} This work was partially funded by the SyncFree Project in the European Union Seventh Framework Programme (FP7/2007-2013) under grant agreement n\textsuperscript{o} 609551, by the LightKone Project in the European Union Horizon 2020 Framework Programme for Research and Innovation (H2020/2014-2020), under grant agreement n\textsuperscript{o} 732505, by SMILES within project ``TEC4Growth – Pervasive Intelligence, Enhancers and Proofs of Concept with Industrial Impact/NORTE-01- 0145-FEDER-000020'' financed by the North Portugal Regional Operational Programme (NORTE 2020), under the PORTUGAL 2020 Partnership Agreement, and through the European Regional Development Fund (ERDF). Chris is funded by the Erasmus Mundus Doctorate Programme under grant agreement n\textsuperscript{o} 2012-0030.}

\balance

\bibliographystyle{ACM-Reference-Format}
\bibliography{lasp}


\begin{thebibliography}{00}


\ifx \showCODEN    \undefined \def \showCODEN     #1{\unskip}     \fi
\ifx \showDOI      \undefined \def \showDOI       #1{#1}\fi
\ifx \showISBNx    \undefined \def \showISBNx     #1{\unskip}     \fi
\ifx \showISBNxiii \undefined \def \showISBNxiii  #1{\unskip}     \fi
\ifx \showISSN     \undefined \def \showISSN      #1{\unskip}     \fi
\ifx \showLCCN     \undefined \def \showLCCN      #1{\unskip}     \fi
\ifx \shownote     \undefined \def \shownote      #1{#1}          \fi
\ifx \showarticletitle \undefined \def \showarticletitle #1{#1}   \fi
\ifx \showURL      \undefined \def \showURL       {\relax}        \fi
\providecommand\bibfield[2]{#2}
\providecommand\bibinfo[2]{#2}
\providecommand\natexlab[1]{#1}
\providecommand\showeprint[2][]{arXiv:#2}

\bibitem[\protect\citeauthoryear{Almeida, Shoker, and Baquero}{Almeida
  et~al\mbox{.}}{2016}]%
        {DBLP:journals/corr/AlmeidaSB16}
\bibfield{author}{\bibinfo{person}{Paulo~S{\'{e}}rgio Almeida},
  \bibinfo{person}{Ali Shoker}, {and} \bibinfo{person}{Carlos Baquero}.}
  \bibinfo{year}{2016}\natexlab{}.
\newblock \showarticletitle{Delta State Replicated Data Types}.
\newblock \bibinfo{journal}{{\em CoRR\/}}  \bibinfo{volume}{abs/1603.01529}
  (\bibinfo{year}{2016}).
\newblock
\showURL{%
\url{http://arxiv.org/abs/1603.01529}}


\bibitem[\protect\citeauthoryear{Alvaro, Conway, Hellerstein, and
  Marczak}{Alvaro et~al\mbox{.}}{2011}]%
        {alvaro2011consistency}
\bibfield{author}{\bibinfo{person}{Peter Alvaro}, \bibinfo{person}{Neil
  Conway}, \bibinfo{person}{Joseph~M Hellerstein}, {and}
  \bibinfo{person}{William~R Marczak}.} \bibinfo{year}{2011}\natexlab{}.
\newblock \showarticletitle{Consistency Analysis in Bloom: a CALM and Collected
  Approach.}. In \bibinfo{booktitle}{{\em CIDR}}. \bibinfo{pages}{249--260}.
\newblock


\bibitem[\protect\citeauthoryear{Burckhardt, Leijen, Protzenko, and
  F{\"a}hndrich}{Burckhardt et~al\mbox{.}}{2015}]%
        {burckhardt2015global}
\bibfield{author}{\bibinfo{person}{Sebastian Burckhardt}, \bibinfo{person}{Daan
  Leijen}, \bibinfo{person}{Jonathan Protzenko}, {and} \bibinfo{person}{Manuel
  F{\"a}hndrich}.} \bibinfo{year}{2015}\natexlab{}.
\newblock \showarticletitle{Global sequence protocol: A robust abstraction for
  replicated shared state}. In \bibinfo{booktitle}{{\em LIPIcs-Leibniz
  International Proceedings in Informatics}}, Vol.~\bibinfo{volume}{37}.
  Schloss Dagstuhl-Leibniz-Zentrum fuer Informatik.
\newblock


\bibitem[\protect\citeauthoryear{{Christopher S. Meiklejohn}}{{Christopher S.
  Meiklejohn}}{2017a}]%
        {lasp-documentation}
\bibfield{author}{\bibinfo{person}{{Christopher S. Meiklejohn}}.}
  \bibinfo{year}{2017}\natexlab{a}.
\newblock \bibinfo{title}{{Lasp Language Documentation}}.
\newblock \bibinfo{howpublished}{\url{https://lasp-lang.org}}.
  (\bibinfo{year}{2017}).
\newblock


\bibitem[\protect\citeauthoryear{{Christopher S. Meiklejohn}}{{Christopher S.
  Meiklejohn}}{2017b}]%
        {lasp-implementation}
\bibfield{author}{\bibinfo{person}{{Christopher S. Meiklejohn}}.}
  \bibinfo{year}{2017}\natexlab{b}.
\newblock \bibinfo{title}{{Lasp Language Source Repository}}.
\newblock \bibinfo{howpublished}{\url{https://github.com/lasp-lang}}.
  (\bibinfo{year}{2017}).
\newblock


\bibitem[\protect\citeauthoryear{Conway, Marczak, Alvaro, Hellerstein, and
  Maier}{Conway et~al\mbox{.}}{2012}]%
        {conway2012logic}
\bibfield{author}{\bibinfo{person}{Neil Conway}, \bibinfo{person}{William~R
  Marczak}, \bibinfo{person}{Peter Alvaro}, \bibinfo{person}{Joseph~M
  Hellerstein}, {and} \bibinfo{person}{David Maier}.}
  \bibinfo{year}{2012}\natexlab{}.
\newblock \showarticletitle{Logic and lattices for distributed programming}. In
  \bibinfo{booktitle}{{\em Proceedings of the Third ACM Symposium on Cloud
  Computing}}. ACM, \bibinfo{pages}{1}.
\newblock


\bibitem[\protect\citeauthoryear{Hindman, Konwinski, Zaharia, Ghodsi, Joseph,
  Katz, Shenker, and Stoica}{Hindman et~al\mbox{.}}{2011}]%
        {hindman2011mesos}
\bibfield{author}{\bibinfo{person}{Benjamin Hindman}, \bibinfo{person}{Andy
  Konwinski}, \bibinfo{person}{Matei Zaharia}, \bibinfo{person}{Ali Ghodsi},
  \bibinfo{person}{Anthony~D Joseph}, \bibinfo{person}{Randy~H Katz},
  \bibinfo{person}{Scott Shenker}, {and} \bibinfo{person}{Ion Stoica}.}
  \bibinfo{year}{2011}\natexlab{}.
\newblock \showarticletitle{Mesos: A Platform for Fine-Grained Resource Sharing
  in the Data Center.}. In \bibinfo{booktitle}{{\em NSDI}},
  Vol.~\bibinfo{volume}{11}. \bibinfo{pages}{22--22}.
\newblock


\bibitem[\protect\citeauthoryear{Kuper and Newton}{Kuper and Newton}{2014}]%
        {kuperjoining}
\bibfield{author}{\bibinfo{person}{Lindsey Kuper} {and} \bibinfo{person}{Ryan~R
  Newton}.} \bibinfo{year}{2014}\natexlab{}.
\newblock \showarticletitle{Joining forces: toward a unified account of LVars
  and convergent replicated data types}. In \bibinfo{booktitle}{{\em 5th
  Workshop on Determinism and Correctness in Parallel Programming (WoDet
  2014)}}.
\newblock


\bibitem[\protect\citeauthoryear{Leit{\~a}o, Pereira, and Rodrigues}{Leit{\~a}o
  et~al\mbox{.}}{2007a}]%
        {leitao2007epidemic}
\bibfield{author}{\bibinfo{person}{Jo{\~a}o Leit{\~a}o}, \bibinfo{person}{Jose
  Pereira}, {and} \bibinfo{person}{Lu{\`\i}s Rodrigues}.}
  \bibinfo{year}{2007}\natexlab{a}.
\newblock \showarticletitle{Epidemic broadcast trees}. IEEE,
  \bibinfo{pages}{301--310}.
\newblock


\bibitem[\protect\citeauthoryear{Leit{\~a}o, Pereira, and Rodrigues}{Leit{\~a}o
  et~al\mbox{.}}{2007b}]%
        {leitao2007hyparview}
\bibfield{author}{\bibinfo{person}{Jo{\~a}o Leit{\~a}o},
  \bibinfo{person}{Jos{\'e} Pereira}, {and} \bibinfo{person}{Lu{\`\i}s
  Rodrigues}.} \bibinfo{year}{2007}\natexlab{b}.
\newblock \showarticletitle{{H}y{P}ar{V}iew: A membership protocol for reliable
  gossip-based broadcast}. IEEE, \bibinfo{pages}{419--429}.
\newblock


\bibitem[\protect\citeauthoryear{Meiklejohn and Van~Roy}{Meiklejohn and
  Van~Roy}{2015a}]%
        {meiklejohn2015lasp}
\bibfield{author}{\bibinfo{person}{Christopher Meiklejohn} {and}
  \bibinfo{person}{Peter Van~Roy}.} \bibinfo{year}{2015}\natexlab{a}.
\newblock \showarticletitle{{Lasp: A language for distributed,
  coordination-free programming}}. In \bibinfo{booktitle}{{\em Proceedings of
  the 17th International Symposium on Principles and Practice of Declarative
  Programming}}. ACM, \bibinfo{pages}{184--195}.
\newblock


\bibitem[\protect\citeauthoryear{Meiklejohn and Van~Roy}{Meiklejohn and
  Van~Roy}{2015b}]%
        {meiklejohn2015selective}
\bibfield{author}{\bibinfo{person}{Christopher Meiklejohn} {and}
  \bibinfo{person}{Peter Van~Roy}.} \bibinfo{year}{2015}\natexlab{b}.
\newblock \showarticletitle{Selective Hearing: An Approach to Distributed,
  Eventually Consistent Edge Computation}. In \bibinfo{booktitle}{{\em Reliable
  Distributed Systems Workshop (SRDSW), 2015 IEEE 34th Symposium on}}. IEEE,
  \bibinfo{pages}{62--67}.
\newblock


\bibitem[\protect\citeauthoryear{Meiklejohn and Van~Roy}{Meiklejohn and
  Van~Roy}{2017}]%
        {meiklejohn2017loquat}
\bibfield{author}{\bibinfo{person}{Christopher~S Meiklejohn} {and}
  \bibinfo{person}{Peter Van~Roy}.} \bibinfo{year}{2017}\natexlab{}.
\newblock \showarticletitle{Loquat: A framework for large-scale actor
  communication on edge networks}. In \bibinfo{booktitle}{{\em Pervasive
  Computing and Communications Workshops (PerCom Workshops), 2017 IEEE
  International Conference on}}. IEEE, \bibinfo{pages}{563--568}.
\newblock


\bibitem[\protect\citeauthoryear{Myter, Coppieters, Scholliers, and
  De~Meuter}{Myter et~al\mbox{.}}{2016}]%
        {myter2016now}
\bibfield{author}{\bibinfo{person}{Florian Myter}, \bibinfo{person}{Tim
  Coppieters}, \bibinfo{person}{Christophe Scholliers}, {and}
  \bibinfo{person}{Wolfgang De~Meuter}.} \bibinfo{year}{2016}\natexlab{}.
\newblock \showarticletitle{I now pronounce you reactive and consistent:
  handling distributed and replicated state in reactive programming}. In
  \bibinfo{booktitle}{{\em Proceedings of the 3rd International Workshop on
  Reactive and Event-Based Languages and Systems}}. ACM, \bibinfo{pages}{1--8}.
\newblock


\bibitem[\protect\citeauthoryear{Rodrigues and Druschel}{Rodrigues and
  Druschel}{2010}]%
        {rodrigues2010peer}
\bibfield{author}{\bibinfo{person}{Rodrigo Rodrigues} {and}
  \bibinfo{person}{Peter Druschel}.} \bibinfo{year}{2010}\natexlab{}.
\newblock \showarticletitle{Peer-to-peer systems}.
\newblock \bibinfo{journal}{{\it Commun. ACM}} \bibinfo{volume}{53},
  \bibinfo{number}{10} (\bibinfo{year}{2010}), \bibinfo{pages}{72--82}.
\newblock


\bibitem[\protect\citeauthoryear{Shapiro, Pregui{\c{c}}a, Baquero, and
  Zawirski}{Shapiro et~al\mbox{.}}{2011}]%
        {shapiro2011comprehensive}
\bibfield{author}{\bibinfo{person}{Marc Shapiro}, \bibinfo{person}{Nuno
  Pregui{\c{c}}a}, \bibinfo{person}{Carlos Baquero}, {and}
  \bibinfo{person}{Marek Zawirski}.} \bibinfo{year}{2011}\natexlab{}.
\newblock \bibinfo{booktitle}{{\em {A comprehensive study of convergent and
  commutative replicated data types}}}.
\newblock \bibinfo{type}{{T}echnical {R}eport} RR-7506.
  \bibinfo{institution}{INRIA}.
\newblock


\bibitem[\protect\citeauthoryear{Sivaramakrishnan, Kaki, and
  Jagannathan}{Sivaramakrishnan et~al\mbox{.}}{2015}]%
        {Sivaramakrishnan:2015:DPO:2737924.2737981}
\bibfield{author}{\bibinfo{person}{KC Sivaramakrishnan},
  \bibinfo{person}{Gowtham Kaki}, {and} \bibinfo{person}{Suresh Jagannathan}.}
  \bibinfo{year}{2015}\natexlab{}.
\newblock \showarticletitle{Declarative Programming over Eventually Consistent
  Data Stores}. In \bibinfo{booktitle}{{\em Proceedings of the 36th ACM SIGPLAN
  Conference on Programming Language Design and Implementation}} {\em
  (\bibinfo{series}{PLDI '15})}. \bibinfo{publisher}{ACM},
  \bibinfo{address}{New York, NY, USA}, \bibinfo{pages}{413--424}.
\newblock
\showISBNx{978-1-4503-3468-6}
\showDOI{%
\url{https://doi.org/10.1145/2737924.2737981}}


\bibitem[\protect\citeauthoryear{Terry, Theimer, Petersen, Demers, Spreitzer,
  and Hauser}{Terry et~al\mbox{.}}{1995}]%
        {terry1995managing}
\bibfield{author}{\bibinfo{person}{Douglas~B Terry}, \bibinfo{person}{Marvin~M
  Theimer}, \bibinfo{person}{Karin Petersen}, \bibinfo{person}{Alan~J Demers},
  \bibinfo{person}{Mike~J Spreitzer}, {and} \bibinfo{person}{Carl~H Hauser}.}
  \bibinfo{year}{1995}\natexlab{}.
\newblock \bibinfo{booktitle}{{\em Managing update conflicts in Bayou, a weakly
  connected replicated storage system}}. Vol.~\bibinfo{volume}{29}.
\newblock \bibinfo{publisher}{ACM}.
\newblock


\end{thebibliography}

\end{document}